\documentclass[12pt]{article}
\usepackage{amsmath,amssymb,bm,graphicx}
\usepackage{color} 

\newcommand{\delslash}{\not \! \partial}

\begin{document}


\begin{center}
{\Large{\bf Majorana neutrino and the vacuum of Bogoliubov quasiparticle}}
\end{center}\vskip .5 truecm
\begin{center}
{\bf { Kazuo Fujikawa
}}
\end{center}

\begin{center}
\vspace*{0.4cm} 
{\it {
Quantum Hadron Physics Laboratory, RIKEN Nishina Center,\\
Wako 351-0198, Japan
}}
\end{center}

\vspace*{0.4cm} 
\begin{abstract} 
The Lagrangian of the seesaw mechanism is C violating but the same Lagrangian when re-written in terms of Majorana neutrinos is manifestly C invariant. To resolve this puzzling feature, a relativistic analogue of Bogoliubov transformation, which preserves CP but explicitly breaks C and P separately, was introduced together with the notions of a Bogoliubov quasiparticle and an analogue of the energy gap in BCS theory.  The idea of Majorana neutrino as Bogoliubov quasiparticle was then suggested. In this paper, we study the vacuum structure of the Bogoliubov quasiparticle which becomes heavy by absorbing the C-breaking.
By treating an infinitesimally small C violating term as an  analogue of the chiral symmetry breaking nucleon mass in the model of Nambu and Jona-Lasinio,  we construct an explicit form of the vacuum of the Bogoliubov quasiparticle which defines Majorana neutrinos in seesaw mechanism.  
The vacuum of the Bogoliubov quasiparticle thus constructed has an analogous condensate structure as the vacuum of the quasiparticle (nucleon) in the Nambu--Jona-Lasinio model.
\end{abstract}

\section{Introduction}

It is known that an effective hermitian Lagrangian, which is analogous to BCS theory,
\begin{eqnarray}\label{1}
{\cal L}&=&\overline{\nu}(x)i\gamma^{\mu}\partial_{\mu}\nu(x) - m\overline{\nu}(x)\nu(x)\nonumber\\
&-&\frac{1}{4}\epsilon_{1}[e^{i\alpha}\nu^{T}(x)C\nu(x) + e^{-i\alpha}\overline{\nu}(x)C\overline{\nu}^{T}(x)]\nonumber\\
&-&\frac{1}{4}\epsilon_{5}[e^{i\beta}\nu^{T}(x)C\gamma_{5}\nu(x) - e^{-i\beta}\overline{\nu}(x)C\gamma_{5}\overline{\nu}^{T}(x)],
\end{eqnarray}
with real parameters $m$, $\epsilon_{1}$, $\epsilon_{5}$,  $\alpha$ and $\beta$ describes a wide variety of fundamental problems such as the seesaw mechanism for neutrino masses~\cite{minkowski, yanagida, mohapatra, fukugita, giunti, bilenky, valle} and the neutron-antineutron oscillations~\cite{kuzmin, mohapatra2, phillips, chang, FT1}.  The fermion field $\nu(x)$ is a Dirac-type four component 
object, 
\begin{eqnarray}
\nu(x)=\nu_{L}(x)+\nu_{R}(x)
\end{eqnarray}
and we should identify $\nu(x)=n(x)$ when we analyze neutron oscillations.  By adjusting the phase freedom of the field $\nu(x)$ we fix the phase convention of charge conjugation as $\nu^{c}(x)=C\overline{\nu(x)}^{T}$, with $C=i\gamma^{2}\gamma^{0}$. The parity is defined by $i\gamma^{0}$ parity, namely, $\nu^{p}(t,\vec{x})=i\gamma^{0}\nu(t,-\vec{x})$, to be consistent with the Majorana condition $\nu(x)=C\overline{\nu(x)}^{T}$, which is essential to analyze the Majorana fermions as in the present paper. Our notational conventions follow Ref.\cite{bjorken} and are briefly summarized in Appendix \ref{AppA}.  It is important that  both our charge conjugation convention and the $i\gamma^{0}$-parity preserve the reality of the Majorana fermion in the Majorana representation. The parameters $\alpha$ and $\beta$ break CP symmetry
and we set $\alpha=\beta=0$ in the present paper, for simplicity.  (One may first set $\beta=0$ in \eqref{1} by using the phase freedom of $\nu(x)$ and then fix C and P operations.  The procedure we described above is more general in fixing the phase freedom of C  operation.) 
The  term with real $\epsilon_{1}\equiv m_{R}+m_{L}$ in the seesaw mechanism preserves $i\gamma^{0}$-parity while the second term with real $\epsilon_{5}\equiv m_{R}-m_{L}$ breaks $i\gamma^{0}$-parity.  The term with $\epsilon_{1}$ thus  preserves charge conjugation symmetry while the term with $\epsilon_{5}$ {\em breaks charge conjugation symmetry}~\cite{FT2}.

If one defines the charge conjugation operator ${\cal C}_{\nu}$ for the neutrino field by
\begin{eqnarray}\label{original-C}
{\cal C}_{\nu}\nu(x){\cal C}_{\nu}^{\dagger} = C\overline{\nu(x)}^{T}=\nu^{c}(x),
\end{eqnarray}
the charge conjugation ${\cal C}_{\nu}$ is broken by the term with $\epsilon_{5}$ in the original Lagrangian \eqref{1}. 
Thus, any mass eigenstates obtained by  diagonalizing the total Lagrangian cannot be the precise eigenstates of C, therefore, they cannot be genuine Majorana neutrinos. 
On the other hand, Lagrangian \eqref{1}, when re-written in terms of Majorana neutrinos, becomes manifestly C invariant since Majorana neutrinos are by definition the precise eigenstates of C. To understand this puzzling feature, we have recently introduced  a relativistic  analogue of Bogoliubov transformation, $(\nu, \nu^{c})\rightarrow (N, N^{c})$, defined as~\cite{FT2, FT3, FT4}
\begin{eqnarray}\label{Bogoliubov}
\left(\begin{array}{c}
            N(x)\\
            N^{c}(x)
            \end{array}\right)
&=& \left(\begin{array}{c}
            \cos\theta\, \nu(x)-\gamma_{5}\sin\theta\, \nu^{c}(x)\\
            \cos\theta\, \nu^{c}(x)+\gamma_{5}\sin\theta\, \nu(x)
            \end{array}\right),
\end{eqnarray}
with
\begin{eqnarray}\label{mixing}
\sin 2\theta =\frac{\epsilon_{5}/2}{\sqrt{m^{2}+(\epsilon_{5}/2)^{2}}}.
\end{eqnarray}
One can confirm the classical consistency condition $N^{c}=C\overline{N}^{T}(x)$ using the expressions of the right-hand side of \eqref{Bogoliubov}.
One can also confirm that the anticommutators are preserved, i.e.,
\begin{eqnarray}\label{anti-comm}
&& \{N(t,\vec{x}), N^{c}(t,\vec{y})\}=\{\nu(t,\vec{x}), \nu^{c}(t,\vec{y})\},\nonumber\\  
&&\{N_{\alpha}(t,\vec{x}), N_{\beta}(t,\vec{y})\}=\{N^{c}_{\alpha}(t,\vec{x}), N^{c}_{\beta}(t,\vec{y})\}=0,
\end{eqnarray}  
using the anticommutation relations of $\nu(t,\vec{x})$ and $\nu^{c}(t,\vec{y})$, and thus the canonicity condition of the Bogoliubov transformation for {\em any time} $t$ is satisfied. 

After the Bogoliubov transformation, the Lagrangian \eqref{1} becomes
\begin{eqnarray}\label{N-field}
{\cal L}&=&\frac{1}{2}\left[\overline{N}(x)\left(i\delslash - M\right)
 N(x)+\overline{N^{c}}(x)\left(i\delslash - M\right)N^{c}(x)\right]\nonumber\\
 &-&\frac{\epsilon_{1}}{4}\left[\overline{N^{c}}(x)N(x) + \overline{N}(x)N^{c}(x) \right],
\end{eqnarray}
where the mass parameter is defined by
\begin{eqnarray}\label{14}
M\equiv \sqrt{m^{2}+(\epsilon_{5}/2)^{2}}.
\end{eqnarray} 
This new Lagrangian \eqref{N-field} preserves both C (now defined $N^{c}=C\overline{N}^{T}$) and $i\gamma^{0}$-parity;
the reason for this change is that  a relativistic  analogue of  Bogoliubov transformation \eqref{Bogoliubov} preserves  CP symmetry but does not preserve the transformation properties under C and $i\gamma^{0}$-parity, separately. For example, the original charge conjugation  $\nu(x) \leftrightarrow \nu^{c}(x)$ in \eqref{Bogoliubov} does not lead to the charge conjugation of the transformed variable
$N(x) \leftrightarrow N^{c}(x)$.   
The C-violating term with $\epsilon_{5}$, which is absorbed into a Dirac mass $M$ of the {\em Bogoliubov quasiparticle} $N(x)$, is analogous to the energy gap in BCS theory.  In this sense, the  quasiparticle $N(x)$ becomes heavy while absorbing the C-breaking. 
Also,  a linear combination of a Dirac fermion and its charge conjugate is mapped to a linear combination of another Dirac fermion and its charge conjugate, and thus the Fock vacuum is expected to be mapped to a new  Fock vacuum. 

When one defines Majorana neutrinos by~\footnote{The definition $\psi_{M}=\frac{1}{\sqrt{2}i}(N(x)-N^{c}(x))$ with an imaginary factor $i$ which satisfies $\psi_{M}=C\overline{\psi_{M}}^{T}$ is often used, but this definition requires an anti-unitary ${\cal C}$ to maintain ${\cal C}\psi_{M}{\cal C}^{\dagger}=\psi_{M}$. 
}
\begin{eqnarray}\label{Majorana-fermion}
\psi_{\pm}(x)=\frac{1}{\sqrt{2}}[N(x)\pm N^{c}(x)],
\end{eqnarray}
one obtains
\begin{eqnarray}\label{Majorana-fermion2}
{\cal L}&=&\frac{1}{2}\{\overline{\psi_{+}}(x)[i\gamma^{\mu}\partial_{\mu}- (M+\epsilon_{1}/2)]\psi_{+}(x) + \overline{\psi_{-}}(x)[i\gamma^{\mu}\partial_{\mu}- (M-\epsilon_{1}/2)]\psi_{-}(x)\}.
\end{eqnarray}
The exact vacuum in the present theory is the Majorana vacuum:
\begin{eqnarray}\label{Majorana-vacuum}
\psi^{(+)}_{+}(x) |0\rangle_{M} = \psi^{(+)}_{-}(x)|0\rangle_{M}=0,
\ \ \ {\cal C}|0\rangle_{M}=|0\rangle_{M},
\end{eqnarray}
where $\psi^{(\pm)}_{+}(x)$ stand for positive frequency components with the charge conjugation properties
\begin{eqnarray}\label{C-of-neutrino}
{\cal C}\psi_{+}(x){\cal C}^{\dagger}=\psi_{+}(x), \ \  {\cal C}\psi_{-}(x){\cal C}^{\dagger}=-\psi_{-}(x), \ \ {\cal C}N(x){\cal C}^{\dagger}=N^{c}(x).
\end{eqnarray}
The smaller mass  (with the common choice $\epsilon_{1}=\epsilon_{5}=m_{R}$)
\begin{eqnarray}\label{neutrino-mass}
m_{\nu}=M-\epsilon_{1}/2=\sqrt{m^{2}+(m_{R}/2)^{2}} - m_{R}/2 \simeq m^{2}/m_{R}
\end{eqnarray}
represents the very small neutrino mass when $m/m_{R}\ll 1$ ({\em seesaw mechanism}).  
It is also confirmed that the Bogoliubov quasiparticle $N(x)$ is defined on the vacuum $|0\rangle_{N}=|0\rangle_{M}$ by noting
\begin{eqnarray}
N(x)=\frac{1}{\sqrt{2}}[\psi_{+}(x)+\psi_{-}(x)], \ \  N^{c}(x)=\frac{1}{\sqrt{2}}[\psi_{+}(x)-\psi_{-}(x)].
\end{eqnarray}
The original field in \eqref{1} is given by 
\begin{eqnarray}\label{original-field}
\left(\begin{array}{c}
            \nu(x)\\
            \nu^{c}(x)
            \end{array}\right)
 &=& \left(\begin{array}{c}
            \cos\theta\, \frac{1}{\sqrt{2}}[\psi_{+}(x)+\psi_{-}(x)]+\gamma_{5}\sin\theta\, \frac{1}{\sqrt{2}}[\psi_{+}(x)-\psi_{-}(x)]\\
            \cos\theta\, \frac{1}{\sqrt{2}}[\psi_{+}(x)-\psi_{-}(x)]-\gamma_{5}\sin\theta\, \frac{1}{\sqrt{2}}[\psi_{+}(x)+\psi_{-}(x)]
            \end{array}\right)           
\end{eqnarray}
and thus  the left-handed neutrino $\nu_{L}(x)$, which describes the weak interaction, and its charge conjugate $\nu_{L}^{c}(x)={\cal C}_{\nu}[(1-\gamma_{5})/2]\nu(x){\cal C}^{\dagger}_{\nu}=[(1-\gamma_{5})/2]\nu^{c}(x)$ are given by
\begin{eqnarray}\label{weak-eigenstate}
\nu_{L}(x)
 &\simeq&(\frac{m}{m_{R}})\psi_{+}(x)_{L}+
 (1-\frac{1}{2}\frac{m^{2}}{m_{R}^{2}})\psi_{-}(x)_{L},\nonumber\\
 \nu_{L}^{c}(x)&\simeq&(1-\frac{1}{2}\frac{m^{2}}{m_{R}^{2}})\psi_{+}(x)_{L}-(\frac{m}{m_{R}})\psi_{-}(x)_{L},       
\end{eqnarray}
using the same notation as in \eqref{neutrino-mass} with 
$m/m_{R}\ll 1$.
Note that  ${\cal C}$ in  \eqref{C-of-neutrino} and ${\cal C}_{\nu}$ in \eqref{original-C} are different, 
${\cal C} \neq {\cal C}_{\nu}$, 
which is seen by recalling that ${\cal C}$ induces $\psi_{\pm}(x) \rightarrow \pm\psi_{\pm}(x)$ while ${\cal C}_{\nu}$ induces  $\nu_{L}(x) \leftrightarrow \nu_{L}^{c}(x)$ in  \eqref{weak-eigenstate}. One recognizes that the physical left-handed neutrino $\nu_{L}(x)$ consists mostly of the left-handed component of the  Majorana neutrino $\psi_{-}(x)$ in \eqref{Majorana-fermion}, but the  deformation of C-symmetry is substantial. This deformation of C-symmetry could become physically  significant if the right-handed neutrino mass is not very large, as is the case in some versions of seesaw mechanism.    

We emphasize that the Lagrangian for the Majorana neutrino \eqref{Majorana-fermion2} with the mass eigenvalues such as \eqref{neutrino-mass} are in agreement with
the common analyses of seesaw mechanism~\cite{minkowski, yanagida, mohapatra, fukugita, giunti, bilenky, valle}, but the charge conjugation is very different since $(\nu_{L}(x))^{c}=C\overline{\nu_{L}(x)}^{T}$ and $(\nu_{R}(x))^{c}=C\overline{\nu_{R}(x)}^{T}$, which make the C-breaking in \eqref{1} unrecognized,  
are used there. The problematic aspects of this C transformation have been clarified in~\cite{FT2,FT3}; for example,  $(\nu_{L}(x))^{c}={\cal C}[(1-\gamma_{5})/2]\nu_{L}(x){\cal C}^{\dagger}=[(1-\gamma_{5})/2]{\cal C}\nu_{L}(x){\cal C}^{\dagger}=[(1-\gamma_{5})/2]C\overline{\nu_{L}(x)}^{T}=0$ and thus logically inconsistent if one assumes a unitary operator which satisfies $(\nu_{L}(x))^{c}={\cal C}\nu_{L}(x){\cal C}^{\dagger}=C\overline{\nu_{L}(x)}^{T}$. The extension of our analysis to the seesaw mechanism of full three generations of neutrinos has been discussed in \cite{FT4}.

The seesaw Lagrangian as formulated in the present paper has no
direct connection with spontaneous breaking of any symmetries. 
Nevertheless, we encountered the interesting analogues of 
a Bogoliubov quasiparticle and the energy-gap in BCS theory, together with a change of the vacuum in the course of our analysis of Majorana neutrinos. Since the Majorana neutrinos are defined in terms of $N(x)$, we  suggested the idea of 
Majorana neutrino as Bogoliubov quasiparticle~\cite{FT3}. 
 In the phenomenology of seesaw mechanism, the definition of the vacuum $\psi^{(+)}_{+}(x) |0\rangle_{M} = \psi^{(+)}_{-}(x)|0\rangle_{M}=0$ is sufficient. But when one identifies $N(x)$ as a Bogoliubov quasiparticle, it would be  interesting to see how the vacuum $|0\rangle_{N}$ of the Bogoliubov quasiparticle, which is precisely defined with $\epsilon_{1}=0$ in \eqref{1},  differs from the naive  vacuum of the Dirac neutrino $|0\rangle_{(0)}$ with $\epsilon_{1}=\epsilon_{5}=0$ in \eqref{1}.  We discuss this problem by following the analysis of Nambu and Jona-Lasinio~\cite{nambu}.  We also clarify the difference of our relativistic analogue of Bogoliubov transformation and the conventional Bogoliubov transformation which is intrinsically non-relativistic.

\section{Conventional Bogoliubov transformation}

In the course of the analysis of spontaneous chiral symmetry breaking~\cite{nambu}, Nambu and Jona-Lasinio  formulated a conventional Bogoliubov transformation in field theory in a very simple manner, although they did not call it Bogoliubov transformation.  The conventional Bogoliubov transformation as it stands does not carry any essential dynamical information of condensate formation, and in this sense may be called  a kinematical transformation consisting of a change of variables to absorb the induced condensate into the mass of the quasiparticle.
However, it shows the vacuum of the quasiparticle in a very intuitive manner by the condensate in the naive vacuum. Also, combined with the consideration of the possible formation of condensed pairs and energetics, it provides a certain qualitative consistency check of spontaneous symmetry breaking~\cite{nambu}. 

We utilize this latter aspect of the conventional Bogoliubov transformation to see the consistency of our identification of $N(x)$ with an analogue of the Bogoliubov quasiparticle. Our model \eqref{1} as it stands is quadratic in dynamical variables and exactly solvable, and thus no dynamics of spontaneous symmetry breaking is contained. Nevertheless, we are going to show that we learn an interesting structure of the vacuum of the quasiparticle by this analysis. In the course of our discussion, we also clarify the difference of our relativistic analogue of Bogoliubov transformation and the conventional Bogoliubov transformation which is intrinsically non-relativistic.

 Nambu and Jona-Lasinio start with a comparison of the model
\begin{eqnarray}\label{massive-model}
{\cal L}&=&\overline{\psi}(x)i\gamma^{\mu}\partial_{\mu}\psi(x) - m\overline{\psi}(x)\psi(x)
\end{eqnarray}
with the ``free'' Lagrangian
\begin{eqnarray}\label{massless-model}
{\cal L}_{0}&=&
\overline{\psi}(x)i\gamma^{\mu}\partial_{\mu}\psi(x). 
\end{eqnarray}
The Lagrangian ${\cal L}_{0}$ is invariant under the conventional  continuous chiral symmetry, but 
the Lagrangian ${\cal L}$ is not invariant under the chiral symmetry due to the finite nucleon mass $m\neq 0$. The spontaneous generation of a finite mass is the main subject of the paper~\cite{nambu}, but we do not discuss this aspect here.

The canonical transformation (Bogoliubov transformation) in~\cite{nambu} is defined for the exact solution $\psi(x)$ of ${\cal L}$ and the exact solution $\psi_{0}(0,\vec{x})$ of ${\cal L}_{0}$ by 
\begin{eqnarray}\label{connection4}
\psi(0,\vec{x})=\psi_{0}(0,\vec{x}),
\end{eqnarray}
for any $\vec{x}$ at a suitable time $t=0$. One then obviously satisfies the canonicity condition, which is the essence of the Bogoliubov transformation, for example,  
\begin{eqnarray}
\{\psi(0,\vec{x}), \psi^{\dagger}(0,\vec{y})\}=\{\psi_{0}(0,\vec{x}), \psi_{0}^{\dagger}(0,\vec{y})\} =\delta(\vec{x}-\vec{y}).
\end{eqnarray}
Namely, one can evaluate the canonical anticommutator of $\psi(0,\vec{x})$ using the field $\psi_{0}(0,\vec{x})$; the creation and annihilation operators contained in $\psi(0,\vec{x})$, when expressed in term of creation and annihilation operators of $\psi_{0}(0,\vec{x})$ using \eqref{connection4}, satisfy the usual canonical anticommutation relations. 
This Bogoliubov transformation as it stands is not Lorentz invariant  since it is based on the specific time $t=0$, although rotation and translation invariance in $d=3$ is maintained. Also the operation of correlating two fields with different masses is a very crude approximation.  But the essence of this identification is the physical insight one obtains, in particular, into the relation of two vacua before and after the spontaneous symmetry breaking.  Intuitively, the above identification would be more accurate when the symmetry breaking parameter, mass $m$ in the present case, is infinitesimal. We thus mainly concentrate on this limit in the present paper. Our main interest is the structure of the vacuum of a quasiparticle introduced by our relativistic analogue of Bogoliubov transformation.
(A manifestly Lorentz covariant treatment of condensate formation and the fermion mass
is given by the effective theory of Goldstone~\cite{goldstone} and its generalization.)

We thus follow the analysis~\cite{nambu}.  
We define the exact solution of \eqref{massive-model} by noting the phase convention of $e^{i(\frac{1-s}{2})}=(-1)^{(\frac{1-s}{2})}$ of spinor eigenfunctions summarized in Appendix \ref{AppA}, 
\begin{eqnarray}
\psi(x)&=&\int [dp] [a_{m}(\vec{p},s)u(\vec{p},s,m)e^{-ipx}
+ b_{m}^{\dagger}(\vec{p},s) v(\vec{p},s,m)e^{ipx}],\nonumber\\
\psi^{c}(x)&=&\int [dp](-1)^{(\frac{1-s}{2})} [b_{m}(\vec{p},s)u(\vec{p},s,m)e^{-ipx}
+ a_{m}^{\dagger}(\vec{p},s) v(\vec{p},s,m)e^{ipx}],
\end{eqnarray}
with $\int [dp]=\sum_{s=\pm 1}\int\frac{d^{3}p}{(2\pi)^{3}}$, and 
\begin{eqnarray}\label{covariant-NJL}
&&a_{m}(\vec{p},s)|0\rangle_{m} = b_{m}(\vec{p},s)|0\rangle_{m}=0,\nonumber\\ 
&& {\cal C}a_{m}(\vec{p},s){{\cal C}}^{\dagger} = (-1)^{(\frac{1-s}{2})}b_{m}(\vec{p},s),\ \  {\cal C}b^{\dagger}_{m}(\vec{p},s){{\cal C}}^{\dagger} = (-1)^{(\frac{1-s}{2})}a^{\dagger}_{m}(\vec{p},s),
\end{eqnarray}
and 
the exact solution of the massless theory \eqref{massless-model}:
\begin{eqnarray}
\psi_{0}(x)&=&\int [dp] [a_{0}(\vec{p},s)u(\vec{p},s,0)e^{-ipx}
+ b_{0}^{\dagger}(\vec{p},s) v(\vec{p},s,0)e^{ipx}],\nonumber\\
\psi_{0}^{c}(x)&=&\int [dp](-1)^{(\frac{1-s}{2})} [b_{0}(\vec{p},s)u(\vec{p},s,0)e^{-ipx}
+ a_{0}^{\dagger}(\vec{p},s) v(\vec{p},s,0)e^{ipx}],
\end{eqnarray}
with 
\begin{eqnarray}
&&a_{0}(\vec{p},s)|0\rangle_{0} = b_{0}(\vec{p},s)|0\rangle_{0}=0,\nonumber\\
&& {\cal C}_{0}a_{0}(\vec{p},s){{\cal C}_{0}}^{\dagger} = (-1)^{(\frac{1-s}{2})}b_{0}(\vec{p},s),\ \  {\cal C}_{0}b^{\dagger}_{0}(\vec{p},s){{\cal C}_{0}}^{\dagger} = (-1)^{(\frac{1-s}{2})}a^{\dagger}_{0}(\vec{p},s). 
\end{eqnarray}
We then have from the condition \eqref{connection4}
\begin{eqnarray}\label{coefficient-NJL}
a_{m}(\vec{p},s)
&&=\alpha_{1}(m,0)a_{0}(\vec{p},s) 
+ (-1)^{(1+s)/2} \alpha_{2}(m,0)b_{0}^{\dagger}(-\vec{p},-s)
,\nonumber\\
b_{m}(\vec{p},s)
&&=\alpha_{1}(m,0)b_{0}(\vec{p},s) 
+ (-1)^{(1-s)/2}\alpha_{2}(m,0)a_{0}^{\dagger}(-\vec{p},-s),
\end{eqnarray}
using the results of detailed evaluations given in \eqref{operator-transformation} below by setting $\theta=0$ there. 
 Note that $s$ is defined by choosing the direction of $\vec{p}$ as quantization axis. 
The vacuum conditions $a_{m}(\vec{p},s)|0\rangle^{(+)}_{m}=0$
and $b_{m}(\vec{p},s)|0\rangle^{(-)}_{m}=0$, respectively, give
 the vacuum of the quasiparticle (nucleon) for an infinitesimally small $m$ (see Appendix \ref{AppB} for the coefficients in \eqref{coefficient-NJL}),  
\begin{eqnarray}\label{NJL-vacuum}
|0\rangle^{(+)}_{m}&=&\prod_{\vec{p},s}\left[1 -\frac{m}{2\sqrt{\vec{p}^{2}+m^{2}}} a^{\dagger}_{0}(\vec{p},s)(-1)^{(1+s)/2} b_{0}^{\dagger}(-\vec{p},-s) \right]|0\rangle_{0},\nonumber\\
|0\rangle^{(-)}_{m}&=&\prod_{\vec{p},s}\left[1 +\frac{m}{2\sqrt{\vec{p}^{2}+m^{2}}} a^{\dagger}_{0}(\vec{p},s)(-1)^{(1+s)/2} b_{0}^{\dagger}(-\vec{p},-s)\right] |0\rangle_{0},
\end{eqnarray}
and one can confirm after suitable normalization as in \eqref{normalized} below that 
${}^{(+)}_{m}\langle 0|0\rangle^{(-)}_{m}=0.$
One can explicitly check that
\begin{eqnarray}
{\cal C}_{0}|0\rangle^{(+)}_{m}=|0\rangle^{(-)}_{m}, \ \ {\cal P}_{0}|0\rangle^{(-)}_{m}=|0\rangle^{(+)}_{m}, \ \ {\cal C}_{0}{\cal P}_{0}|0\rangle^{(+)}_{m}=|0\rangle^{(+)}_{m},
\end{eqnarray}
where parity is defined by $a^{\dagger}_{0}(\vec{p},s)\rightarrow ia_{0}(-\vec{p},s)$ and $b_{0}(\vec{p},s) \rightarrow ib_{0}(-\vec{p},s)$ in the present $i\gamma^{0}$-parity.  We thus choose  
\begin{eqnarray}
{\cal C}{\cal P}|0\rangle^{(+)}_{m}={\cal C}_{0}{\cal P}_{0}|0\rangle^{(+)}_{m}=|0\rangle^{(+)}_{m},
\end{eqnarray} 
which shows that CP invariant theory, as in the present example, is defined solely on the vacuum 
$|0\rangle^{(+)}_{m}$.  The appearance of the degenerate and orthogonal vacua is a characteristic of the present formulation, which might be an artifact of non-relativistic treatment such as static pair condensation.  In comparison, 
the direct equivalence $|0\rangle^{(+)}_{m}=|0\rangle^{(-)}_{m}$ was used by a phase choice in the original paper~\cite{nambu}. Physically, the vacuum of the quasiparticle is the condensation of massless nucleon-antinucleon pairs with  $\pm2$ units of chirality~\cite{nambu}.

\section{Vacuum of Bogoliubov quasiparticle}
The analogues of the energy gap and the Bogoliubov quasiparticle appear  when we apply a relativistic analogue of the Bogoliubov transformation to the fermion-number violating Lagrangian ${\cal L}$ in \eqref{1}, with a tentative choice  $\epsilon_{1}=0$  and treating the C-symmetry breaking term with $\epsilon_{5}$ as a possible condensation.  The C-invariant term with $\epsilon_{1}\neq 0$ then induces an enormous mass splitting of these degenerate Majorana fermions. 

To study the vacuum of the Bogoliubov quasiparticle $N(x)$, 
we thus start with the Lagrangian which is obtained by tentatively setting $\epsilon_{1}=0$,
\begin{eqnarray}\label{simplified-model1}
{\cal L}&=&\overline{\nu}(x)i\gamma^{\mu}\partial_{\mu}\nu(x) - m\overline{\nu}(x)\nu(x)\nonumber\\
&-&\frac{1}{4}\epsilon_{5}[\nu^{T}(x)C\gamma_{5}\nu(x) - \overline{\nu}(x)C\gamma_{5}\overline{\nu}^{T}(x)],
\end{eqnarray}
and we later discuss how to incorporate the effects of the term with $\epsilon_{1}$. We thus analyze the model \eqref{1} in the parameter space $(\epsilon_{1}, \epsilon_{5})$ by looking at specific corners. This Lagrangian is re-written for the quasiparticle with mass $M$ 
by our Bogoliubov transformation \eqref{Bogoliubov} as
\begin{eqnarray}\label{exact-N}
{\cal L}&=&\overline{N}(x)i\gamma^{\mu}\partial_{\mu}N(x) 
- M \overline{N} N,
\end{eqnarray}
with $M=\sqrt{m^{2}+(\epsilon_{5}/2)^{2}}$. Note that $\epsilon_{5}$ is absorbed into the mass term $M$, namely, it behaves precisely like the energy gap in BCS theory. This ${\cal L}$ is invariant under the charge conjugation symmetry  defined by 
\begin{eqnarray}
&&{\cal C} N(x){\cal C}^{\dagger}=C\overline{N(x)}^{T} =N^{c}(x), \ \ {\cal C}|0\rangle_{N}=|0\rangle_{N}.
\end{eqnarray}
This vacuum is also the vacuum of the exact solution $\nu(x)$ in \eqref{simplified-model1}.

We consider another Lagrangian obtained by setting $\epsilon_{5}=0$ in \eqref{simplified-model1}:
\begin{eqnarray}\label{new-L}
{\cal L}_{0}&=&
\overline{\nu}(x)i\gamma^{\mu}\partial_{\mu}\nu(x) - m\overline{\nu}(x)\nu(x),
\end{eqnarray}
which is invariant under the C-symmetry defined by 
\begin{eqnarray}
{\cal C}_{\nu} \nu(x){{\cal C}_{\nu}}^{\dagger}=C\overline{\nu(x)}^{T} =\nu^{c}(x), \ \  {\cal C}_{\nu}|0\rangle_{(0)}=|0\rangle_{(0)}.
\end{eqnarray}
This Lagrangian describes a Dirac neutrino.
In both cases, the Dirac-type fermions with definite fermion number and C-symmetry are the exact mass eigenstates. The C-symmetry generated by ${\cal C}_{\nu} $ is broken by the term with $\epsilon_{5}\neq 0$ in \eqref{simplified-model1}. 

We thus recognize the term with $\epsilon_{5}$ as an analogue of the chiral symmetry breaking mass term of Nambu and Jona-Lasinio (but  in their model, there is no analogue of $C$ of $N(x)$).        
To define the conventional Bogoliubov transformation, we expand the exact solution of \eqref{exact-N},
\begin{eqnarray}\label{N-operator}
N(t,\vec{x})&=&\int [dp] [a_{N}(\vec{p},s,M) u(\vec{p},s,M)e^{-i px} + b_{N}^{\dagger}(\vec{p},s,M) v(\vec{p},s,M)e^{ipx}],\\
N^{c}(t,\vec{x})&=&\int [dp] e^{i(\frac{1-s}{2})} [b_{N}(\vec{p},s,M)u(\vec{p},s,M)e^{-ipx} + a_{N}^{\dagger}(\vec{p},s,M)v(\vec{p},s,M)e^{ipx}]\nonumber
\end{eqnarray}
with
\begin{eqnarray}
&&a_{N}(\vec{p},s,M)|0\rangle_{N} = b_{N}(\vec{p},s,M)|0\rangle_{N}=0,\ \ \  
{\cal C}|0\rangle_{N} = |0\rangle_{N}, \\
&& {\cal C}a_{N}(\vec{p},s,M){\cal C}^{\dagger} = e^{i(\frac{1-s}{2})}b_{N}(\vec{p},s,M),\ \  {\cal C}b^{\dagger}_{N}(\vec{p},s,M){\cal C}^{\dagger} = e^{i(\frac{1-s}{2})}a^{\dagger}_{N}(\vec{p},s,M), \nonumber
\end{eqnarray}
and the exact solution of \eqref{new-L}, which we denote by $\nu_{0}(x)$,
\begin{eqnarray}\label{free}
\nu_{0}(x)&=&\int [dp] [a_{\nu}(\vec{p},s,m)u(\vec{p},s,m)e^{-ipx}
+ b_{\nu}^{\dagger}(\vec{p},s,m) v(\vec{p},s,m)e^{ipx}],\nonumber\\
\nu_{0}^{c}(x)&=&\int [dp]e^{i(\frac{1-s}{2})} [b_{\nu}(\vec{p},s,m)u(\vec{p},s,m)e^{-ipx}
+ a_{\nu}^{\dagger}(\vec{p},s,m) v(\vec{p},s,m)e^{ipx}],
\end{eqnarray}
with
\begin{eqnarray}\label{C-symmetry-Dirac}
&&a_{\nu}(\vec{p},s,m)|0\rangle_{(0)} = b_{\nu}(\vec{p},s,m)|0\rangle_{(0)}=0,\ \  
{\cal C}_{\nu}|0\rangle_{(0)} = |0\rangle_{(0)},\\
&& {\cal C}_{\nu}a_{\nu}(\vec{p},s,m){{\cal C}_{\nu}}^{\dagger} = e^{i(\frac{1-s}{2})}b_{\nu}(\vec{p},s,m),\ \  {\cal C}_{\nu}b^{\dagger}_{\nu}(\vec{p},s,m){{\cal C}_{\nu}}^{\dagger} = e^{i(\frac{1-s}{2})}a^{\dagger}_{\nu}(\vec{p},s,m).\nonumber 
\end{eqnarray}
The extra phase factor
\begin{eqnarray}
e^{i(\frac{1-s}{2})}=(-1)^{(\frac{1-s}{2})}
\end{eqnarray}  
arises from the definition of the spinor eigenfunctions in \cite{bjorken} summarized in Appendix \ref{AppA}.
One may then define a canonical transformation at $t=0$ between the exact solution $\nu(x)$ of ${\cal L}$ \eqref{simplified-model1} and the exact solution $\nu_{0}(x)$ of ${\cal L}_{0}$ \eqref{new-L},
which is  the conventional Bogoliubov transformation as formulated by Nambu and Jona-Lasinio, $\nu_{0}(0,\vec{x})=\nu(0,\vec{x})$ and $\nu_{0}^{c}(0, \vec{x})=\nu^{c}(0,\vec{x})$. Apparently, $\nu_{0}(0,\vec{x})$ and $\nu(0,\vec{x})$ thus defined satisfy the same canonical commutation relations at $t=0$.  Recalling our Bogoliubov transformation \eqref{Bogoliubov}, we have
\begin{eqnarray}\label{connection}
\nu_{0}(0,\vec{x})&=&\nu(0,\vec{x})=[\cos\theta N(0,\vec{x}) + \gamma_{5}\sin\theta N^{c}(0,\vec{x})],\nonumber\\
\nu_{0}^{c}(0, \vec{x})&=&\nu^{c}(0,\vec{x})=[\cos\theta N^{c}(0,\vec{x}) - \gamma_{5} \sin\theta N(0,\vec{x})].
\end{eqnarray}
This  is solved as 
\begin{eqnarray}\label{second-connection}
N(0,\vec{x})=[\cos\theta \nu_{0}(0,\vec{x}) - \gamma_{5}\sin\theta \nu_{0}^{c}(0, \vec{x})],\nonumber\\
N^{c}(0,\vec{x})=[\cos\theta \nu_{0}^{c}(0, \vec{x}) + \gamma_{5} \sin\theta \nu_{0}(0,\vec{x})],
\end{eqnarray}
which is a canonical transformation $(\nu_{0}, \nu^{c}_{0}) \rightarrow (N,N^{c})$, CP preserving but C violation. This canonical transformation is, however, defined only at a very specific time $t=0$. 

We thus have from \eqref{second-connection}, using the operator expansions in \eqref{N-operator} and \eqref{free},
\begin{eqnarray}
&&\sum_{s}[a_{N}(\vec{p},s,M) u(\vec{p},s,M) + b_{N}^{\dagger}(-\vec{p},s,M) v(-\vec{p},s,M)]\nonumber\\
&&=\sum_{s}\cos\theta[a_{\nu}(\vec{p},s,m) u(\vec{p},s,m)
+ b_{\nu}^{\dagger}(-\vec{p},s,m) v(-\vec{p},s,m)]\nonumber\\
&&- \sum_{s}\gamma_{5}\sin\theta e^{i(\frac{1-s}{2})}[b_{\nu}(\vec{p},s,m) u(\vec{p},s,m)
+ a_{\nu}^{\dagger}(-\vec{p},s,m) v(-\vec{p},s,m)],
\end{eqnarray}
and a similar relation  for the second relation in \eqref{second-connection}. These relations  
give, using $u(\vec{p},s,M)^{\dagger}u(\vec{p},s^{\prime},M)=\delta_{s,s^{\prime}}$ and $u(\vec{p},s,M)^{\dagger} v(-\vec{p},s^{\prime},M)=0$, 
\begin{eqnarray}
a_{N}(\vec{p},s,M)
&=&
\sum_{s^{\prime}}\cos\theta\{[u(\vec{p},s,M)^{\dagger}
u(\vec{p},s^{\prime},m)]a_{\nu}(\vec{p},s^{\prime},m)\nonumber\\
&&\hspace{1.5cm}+  [u(\vec{p},s,M)^{\dagger}v(-\vec{p},s^{\prime},m)]b_{\nu}^{\dagger}(-\vec{p},s^{\prime},m)\}\nonumber\\
&-&\sum_{s^{\prime}} \sin\theta e^{i(\frac{1-s^{\prime}}{2})}\{ [u(\vec{p},s,M)^{\dagger} \gamma_{5}u(\vec{p},s^{\prime},m)]b_{\nu}(\vec{p},s^{\prime},m)\nonumber\\
&&\hspace{2.5cm}+[u(\vec{p},s,M)^{\dagger}\gamma_{5}v(-\vec{p},s^{\prime},m)]a_{\nu}^{\dagger}(-\vec{p},s^{\prime},m)\},
\end{eqnarray}
and similarly
\begin{eqnarray}
e^{i(\frac{1-s}{2})}b_{N}(\vec{p},s,M)
&=&\sum_{s^{\prime}}\cos\theta e^{i(\frac{1-s^{\prime}}{2})}\{[u(\vec{p},s,M)^{\dagger} u(\vec{p},s^{\prime},m)]b_{\nu}(\vec{p},s^{\prime},m) \nonumber\\
&&\hspace{2.5cm}+ u(\vec{p},s,M)^{\dagger} v(-\vec{p},s^{\prime},m)]a_{\nu}^{\dagger}(-\vec{p},s^{\prime},m)\} \nonumber\\
&+&\sum_{s^{\prime}}\sin\theta \{[ u(\vec{p},s,M)^{\dagger}\gamma_{5} u(\vec{p},s^{\prime},m)]a_{\nu}(\vec{p},s^{\prime},m)
\nonumber\\
&&\hspace{1.5cm}+  [u(\vec{p},s,M)^{\dagger} \gamma_{5}v(-\vec{p},s^{\prime},m)]b_{\nu}^{\dagger}(-\vec{p},s^{\prime},m)\}.
\end{eqnarray}
We have several coefficients to be evaluated, which are given in Appendix \ref{AppB}.  
We then obtain:
\begin{eqnarray}\label{operator-transformation}
&&a_{N}(\vec{p},s,M)\nonumber\\
&&=[\cos\theta \alpha_{1}(M,m)a_{\nu}(\vec{p},s,m) - \sin\theta\beta_{1}(M,m)(-1)^{\frac{1+s}{2}}a_{\nu}^{\dagger}(-\vec{p},-s,m)]\nonumber\\
&&+ (-1)^{(1+s)/2} [\cos\theta\alpha_{2}(M,m)b_{\nu}^{\dagger}(-\vec{p},-s,m)-\sin\theta\beta_{2}(M,m)(-1)^{(1+s)/2}b_{\nu}(\vec{p},s,m)]
,\nonumber\\
\nonumber\\
&&(-1)^{(1-s)/2}b_{N}(\vec{p},s,M)\nonumber\\
&&=[\cos\theta \alpha_{1}(M,m)(-1)^{(1-s)/2}b_{\nu}(\vec{p},s,m) + \sin\theta\beta_{1}(M,m)b_{\nu}^{\dagger}(-\vec{p},-s,m)]\nonumber\\
&&+ [ \cos\theta\alpha_{2}(M,m)a_{\nu}^{\dagger}(-\vec{p},-s,m)+\sin\theta\beta_{2}(M,m)(-1)^{(1-s)/2}a_{\nu}(\vec{p},s,m)].
\end{eqnarray}
We simplified some of these coefficients by choosing the spin-axis in the direction of $\vec{p}$, namely, helicity, but we still use the conventional spin notation.

\subsection{ Vacuum of the quasiparticle $N(x)$}

We have the vacuum condition for quasiparticles:
\begin{eqnarray}
a_{N}(\vec{p},s,M)|0\rangle_{N}=0,\ \ b_{N}(\vec{p},s,M)|0\rangle_{N}=0.
\end{eqnarray}
To satisfy $a_{N}(\vec{p},s,M)|0\rangle_{N}=0$ starting with the Dirac vacuum $|0\rangle_{(0)}$, 
we choose
\begin{eqnarray}\label{plus}
|0\rangle_{(+)}&=&\prod_{(\vec{p},s)}[1 + \frac{\sin\theta\beta_{1}(M,m)}{\cos\theta \alpha_{1}(M,m)}a^{\dagger}_{\nu}(\vec{p},s,m)(-1)^{\frac{1+s}{2}}a_{\nu}^{\dagger}(-\vec{p},-s,m)]\nonumber\\
&\times&[1 + \frac{\cos\theta \alpha_{2}(M,m)}{\sin\theta\beta_{2}(M,m)}b^{\dagger}_{\nu}(\vec{p},s,m)(-1)^{\frac{1+s}{2}}b_{\nu}^{\dagger}(-\vec{p},-s,m)]|0\rangle_{(0)}.
\end{eqnarray}
We do not normalize the vacuum state for the moment. 
Similarly, the condition $b_{N}(\vec{p},s,M)|0\rangle_{N}=0$ is satisfied by 
\begin{eqnarray}\label{minus}
|0\rangle_{(-)}&=&\prod_{(\vec{p},s)}[1 + \frac{\cos\theta \alpha_{2}(M,m)}{\sin\theta\beta_{2}(M,m)}a^{\dagger}_{\nu}(\vec{p},s,m)(-1)^{\frac{1+s}{2}}a_{\nu}^{\dagger}(-\vec{p},-s,m)]\nonumber\\
&\times&[1 + \frac{\sin\theta\beta_{1}(M,m)}{\cos\theta \alpha_{1}(M,m)}b^{\dagger}_{\nu}(\vec{p},s,m)(-1)^{\frac{1+s}{2}}b_{\nu}^{\dagger}(-\vec{p},-s,m)]|0\rangle_{(0)}.
\end{eqnarray}
In the vacuum \eqref{plus} and \eqref{minus} we have condensation factors which are the superpositions of paired states with fermion number $\pm 2$.

In the limit of the small $\epsilon_{5}$, one obtains a representation of \eqref{plus} and \eqref{minus} analogous to \eqref{NJL-vacuum}. Retaining only terms linear in $\epsilon_{5}$ in the small $\epsilon_{5}$ limit, we have from \eqref{plus} and \eqref{minus} using \eqref{mixing} and the results in Appendix \ref{AppB},
\begin{eqnarray}\label{vacuum-quasiparticle}
|0\rangle_{(+)}&=&\prod_{(\vec{p},s)}[1 + \frac{\epsilon_{5}/2}{2\sqrt{\vec{p}^{2}+m^{2}}}a^{\dagger}_{\nu}(\vec{p},s,m)(-1)^{\frac{1+s}{2}}a_{\nu}^{\dagger}(-\vec{p},-s,m)]\nonumber\\
&\times&[1 + \frac{\epsilon_{5}/2}{2\sqrt{\vec{p}^{2}+m^{2}}}b^{\dagger}_{\nu}(\vec{p},s,m)(-1)^{\frac{1+s}{2}}b_{\nu}^{\dagger}(-\vec{p},-s,m)]|0\rangle_{(0)}\nonumber\\
&=&|0\rangle_{(-)}\nonumber\\
&\equiv&|0\rangle_{N}
\end{eqnarray}
which shows a unique C-invariant vacuum ${\cal C}|0\rangle_{N}=|0\rangle_{N}$ 
in the sense
\begin{eqnarray}
a_{N}(\vec{p},s,M)|0\rangle_{N}=0, \ \ \ b_{N}(\vec{p},s,M)|0\rangle_{N}=0,
\end{eqnarray}
combined with ${\cal C}a_{N}(\vec{p},s,M){\cal C}^{\dagger}=(-1)^{\frac{1-s}{2}}b_{N}(\vec{p},s,M)$. 
One also observes that the vacuum is not invariant under the original C-symmetry in \eqref{C-symmetry-Dirac}:
\begin{eqnarray}\label{vacuum-quasiparticle2}
{\cal C}_{\nu}|0\rangle_{N}&=&\prod_{(\vec{p},s)}[1 - \frac{\epsilon_{5}/2}{2\sqrt{\vec{p}^{2}+m^{2}}}a^{\dagger}_{\nu}(\vec{p},s,m)(-1)^{\frac{1+s}{2}}a_{\nu}^{\dagger}(-\vec{p},-s,m)]\nonumber\\
&\times&[1 - \frac{\epsilon_{5}/2}{2\sqrt{\vec{p}^{2}+m^{2}}}b^{\dagger}_{\nu}(\vec{p},s,m)(-1)^{\frac{1+s}{2}}b_{\nu}^{\dagger}(-\vec{p},-s,m)]|0\rangle_{(0)}\nonumber\\
&\neq&|0\rangle_{N},
\end{eqnarray}
although  $\{{\cal C}_{\nu}\}^{2}|0\rangle_{N}=|0\rangle_{N}$. This is consistent with our expectation
\begin{eqnarray}
{\cal C} \neq {\cal C}_{\nu}.
\end{eqnarray}
This analysis shows that the vacuum $|0\rangle_{N}$ is regarded as a vacuum  of ``condensed C-symmetry'' when seen from the original Dirac vacuum 
$|0\rangle_{(0)}$, and  $|0\rangle_{N}\neq |0\rangle_{(0)}$. As for the doublet representation of ${\cal C}_{\nu}$ on $|0\rangle_{N}$, $\{|0\rangle_{N}, {\cal C}_{\nu}|0\rangle_{N} \}$, it may be a discrete C-symmetry analogue of the degenerate vacua for spontaneously broken continuous symmetry~\cite{nambu}. 

One can also confirm CP symmetry explicitly 
${{\cal C}_{\nu}{\cal P}_{\nu}}|0\rangle_{N}=|0\rangle_{N}$ using\\ ${\cal P}_{\nu}a_{\nu}(\vec{p},s,m){{\cal P}_{\nu}}^{\dagger}= i a_{\nu}(-\vec{p},s,m)$ and ${\cal P}_{\nu}b_{\nu}(\vec{p},s,m){{\cal P}_{\nu}}^{\dagger}= i b_{\nu}(-\vec{p},s,m)$ in the present $i\gamma^{0}$-parity in \eqref{vacuum-quasiparticle2}.  
We also have ${\cal P}|0\rangle_{N}=|0\rangle_{N}$ since $a_{N}(\vec{p},s,M)|0\rangle_{N}=0$ and $b_{N}(\vec{p},s,M)|0\rangle_{N}=0$ imply $ia_{N}(-\vec{p},s,M){\cal P}|0\rangle_{N}=0$ and $ib_{N}(-\vec{p},s,M){\cal P}|0\rangle_{N}=0$, respectively.
We thus have the natural relation ${\cal C}{\cal P}|0\rangle_{N}={{\cal C}_{\nu}{\cal P}_{\nu}}|0\rangle_{N}$.  The vacuum \eqref{vacuum-quasiparticle} gives a very explicit construction of the vacuum of a Bogoliubov quasiparticle $N(x)$ from the original Dirac vacuum, although for a specific value of $\epsilon_{5}$. 

When one normalizes the vacuum state, one has 
\begin{eqnarray}\label{normalized}
|0\rangle_{N}&=&\prod_{(\vec{p},s)}\Big[\frac{1}{\sqrt{1+ (\frac{\epsilon_{5}}{4E(p)})^{2}} }+ \frac{\frac{\epsilon_{5}}{4E(p)}}{\sqrt{1+ (\frac{\epsilon_{5}}{4E(p)})^{2}} }a^{\dagger}_{\nu}(\vec{p},s,m)(-1)^{\frac{1+s}{2}}a_{\nu}^{\dagger}(-\vec{p},-s,m)\Big]\nonumber\\
&\times&\Big[\frac{1}{\sqrt{1+ (\frac{\epsilon_{5}}{4E(p)})^{2}} }+ \frac{\frac{\epsilon_{5}}{4E(p)}}{\sqrt{1+ (\frac{\epsilon_{5}}{4E(p)})^{2}}} b^{\dagger}_{\nu}(\vec{p},s,m)(-1)^{\frac{1+s}{2}}b_{\nu}^{\dagger}(-\vec{p},-s,m)\Big]|0\rangle_{(0)},\nonumber\\
\end{eqnarray}
with $E(p)\equiv \sqrt{\vec{p}^{2}+m^{2}}$. One can then confirm the orthogonality of the vacua
${}_{(0)}\langle 0|0\rangle_{N}=
\prod_{(\vec{p},s)}\frac{1}{1+ (\frac{\epsilon_{5}}{4E(p)})^{2} }
=\exp\left[-2\int \frac{d^{3}p}{(2\pi)^{3}}
\ln \left(1+ (\epsilon_{5}/4E(p))^{2}\right)\right]=\exp[-\infty]=0.$
Note that this divergence in the integral is proportional to $\epsilon^{2}_{5}$, which is analogous to the case of the model of  Nambu and Jona-Lasinio~\cite{nambu}, where the divergence is proportional to the induced chiral symmetry breaking mass $m^{2}$.
Energetically also, the small $\epsilon_{5}$ increases the mass $m\rightarrow \sqrt{m^{2}+(\epsilon_{5}/2)^{2}}$ and is thus favored as in the Nambu--Jona-Lasinio model~\cite{nambu}. 
It is remarkable that 
we recognized the consistency with possible spontaneous C-symmetry breaking for an infinitesimal symmetry breaking parameter using the conventional Bogoliubov transformation, by treating $\epsilon_{5}$ as an analogue of the nucleon mass in the Nambu--Jona-Lasinio model. 
However, 
the vacuum \eqref{normalized} consisting of static pair condensation in the Dirac vacuum, which is based on the consistency analysis using the conventional Bogoliubov transformation in the small $\epsilon_{5}$ limit without any dynamical consideration, should be interpreted only as an indication of the possible form of the vacuum of the quasiparticle $N(x)$.

\section{Discussion and conclusion}

It is perfectly consistent to understand a relativistic analogue of the  Bogoliubov 
transformation \eqref{Bogoliubov} in the seesaw mechanism~\cite{FT2, FT3}  as a special class of the Bogoliubov transformation: Our C-breaking transformation in \eqref{Bogoliubov} as it stands  is manifestly Lorentz invariant and canonical for any time $t$ and thus exact, and no-mixing of creation and annihilation operators is induced although the particle and antiparticle are mixed~\footnote{The mixing of creation and annihilation operators and the mixing of particle and antiparticle are very different. One may regard an inverse of \eqref{Bogoliubov} as a change of dynamical variables $(N(x), N^{c}(x)) \rightarrow (\nu(x), \nu^{c}(x))$ in which the mixing of $N(x)$ and $N^{c}(x)$ naturally appears. But the fact that the transformation is defined for any time $t$ implies that the positive and negative frequency components of $\nu(x)$ and $\nu^{c}(x)$, respectively,  are defined in terms of the
positive and negative frequency components of $N(x)$ and $N^{c}(x)$, and thus no mixing of creation and annihilation operators; this is explicitly checked in the present exactly solvable model. In contrast, for the conventional Bogoliubov transformation defined only at $t=0$ for fields with different masses \eqref{connection}, the mixing of creation and annihilation operators  generally takes place due to the energy uncertainty relation (being defined only at the specific instant $t=0$) and the loss of relativity. 
}. The transformation may be called kinematical in view of the absence of the information of any condensate formation  and no information about the vacuum of the Bogoliubov quasiparticle $N(x)$ except for $a_{N}|0\rangle_{N}=b_{N}|0\rangle_{N}=0$.
 Still, the entire off-diagonal mass term with $\epsilon_{5}$ which breaks C-symmetry is absorbed into the Dirac-type diagonal mass term of the Bogoliubov quasiparticle in \eqref{N-field},
which is an essence of the original Bogoliubov transformation in BCS theory.  The possible dynamical origin of the C-violating term with $\epsilon_{5}$ is beyond the scope of our Bogoliubov transformation.

Nevertheless, the relativistic analogue of Bogoliubov transformation identifies a C-violating term with $\epsilon_{5}$ as an analogue of the energy gap in BCS theory as mentioned above, and thus it would be interesting to 
examine what happens if the C-symmetry breaking term is treated in a manner analogous to the chiral symmetry breaking mass term in the model of Nambu and Jona-Lasinio. (Note that one has the covariant relation $a_{m}(\vec{p},s)|0\rangle_{m} = b_{m}(\vec{p},s)|0\rangle_{m}=0$ in \eqref{covariant-NJL} but the explicit construction \eqref{NJL-vacuum} provides a more
illuminating condensate structure in the model~\cite{nambu}.)
For the case with the parameter $\epsilon_{1}$ tentatively set $\epsilon_{1}= 0$, the Majorana neutrinos are degenerate in mass and 
 we  found a very explicit construction of the vacuum of a Bogoliubov quasiparticle $N(x)$ in terms of 
 the Dirac vacuum as in \eqref{normalized} in the small $\epsilon_{5}$ limit.
 This vacuum of the Bogoliubov quasiparticle has a similar condensate structure as the vacuum of the quasiparticle (nucleon) in the Nambu--Jona-Lasinio model \eqref{NJL-vacuum}.  
This may suggest  that there might be some deep dynamics which triggers  the $\epsilon_{5}$ condensation in the seesaw model, such as some kind of dynamical generation of the grand unification scale. 
Finally,  the term with $\epsilon_{1}$, which preserves C-symmetry but breaks lepton number symmetry,  is turned on and induces a mass splitting of Majorana neutrinos. Quite phenomenologically, this two-step procedure is close to how we diagonalize the seesaw Lagrangian \eqref{1} exactly using a relativistic analogue of the Bogoliubov transformation.

In conclusion, we have analyzed the model \eqref{1} at the corners of the parameter space $(\epsilon_{1}, \epsilon_{5})$. We have identified an explicit form of the vacuum \eqref{normalized}  of the Bogoliubov quasiparticle $N(x)$, which becomes heavy by absorbing the C-breaking,  at the corner close to the Dirac point (namely, both  $\epsilon_{1}$ and  $\epsilon_{5}$ are small), using the conventional Bogoliubov transformation formulated by Nambu and Jona-Lasinio.  The extension of  \eqref{normalized} to the realistic parameter domain of seesaw mechanism is not known, but the existence of \eqref{normalized} which satisfies all the required properties is interesting and suggestive.  
\\

I thank A. Tureanu for numerous helpful comments and M. Chaichian for a critical comment. 
\\

\noindent Note added:\\
Our relativistic analogue of the Bogoliubov transformation, which was introduced to define the proper charge conjugation of the Majorana neutrino starting with the C-violating seesaw Lagrangian, is related to the Pauli-G\"ursey transformation \cite{pauli}\cite{gursey}. 
An analogy of the seesaw mechanism to the Nambu{--}Jona-Lasinio model has been briefly mentioned in the original suggestion of the seesaw mechanism by Yanagida in \cite{yanagida}. I thank T. Yanagida for a helpful discussion.

\appendix

\section{Notational convention}\label{AppA}
We follow the Bjorken-Drell convention~\cite{bjorken}, but our choice of spinor solutions, which include a factor $\sqrt{m/E}$, are given by
\begin{eqnarray}\label{Dirac-eigenfunction}
&&u(\vec{p},s, m)=\sqrt{\frac{E+m}{2E}}\left(\begin{array}{c}
            \xi(s)\\
            \frac{\vec{\sigma}\cdot\vec{p}}{E+m}\xi(s)
            \end{array}\right), \ \ 
v(\vec{p},s, m)=\sqrt{\frac{E+m}{2E}}\left(\begin{array}{c}
            \frac{\vec{\sigma}\cdot\vec{p}}{E+m}\xi(-s)\\
            \xi(-s)
            \end{array}\right)\nonumber
\end{eqnarray}
with a two-component spinor $\xi(\pm 1)$ defined at the rest frame.
We normalize the phase convention of the solutions of Dirac equation  to satisfy
\begin{eqnarray}
(-1)^{(\frac{1-s}{2})}v(\vec{p},s,m) = C\overline{u(\vec{p},s,m)}^{T}
\end{eqnarray}
and thus $(-1)^{(\frac{1-s}{2})}u(\vec{p},s,m) = C\overline{v(\vec{p},s,m)}^{T}$ with $C=i\gamma^{2}\gamma^{0}$. When one chooses the direction of $\vec{p}$ as spin quantization, $s$ agrees with helicity which was used in~\cite{nambu}. We have 
orthogonality relations
\begin{eqnarray}
u(\vec{p},s)^{\dagger}u(\vec{p},s^{\prime})=\delta_{s,s^{\prime}}, \ \ \
v(\vec{p},s)^{\dagger}v(\vec{p},s^{\prime})
=\delta_{s,s^{\prime}}, \ \ \
u(\vec{p},s)^{\dagger}v(-\vec{p},s^{\prime})=0.
\end{eqnarray}

\section{Evaluation of transformation coefficients}\label{AppB}
 Using the convention  in Appendix A with $E=\sqrt{\vec{p}^{2}+M^{2}}$ and 
$E^{\prime}=\sqrt{\vec{p}^{2}+m^{2}}$,
\begin{eqnarray}
&&u(\vec{p},s,M)^{\dagger}u(\vec{p},s^{\prime},m)\nonumber\\
&&=\delta_{s,s^{\prime}}\frac{1}{2\sqrt{EE^{\prime}}}[\sqrt{(E+M)(E^{\prime}+m)} +\sqrt{(E-M)(E^{\prime}-m)}]\nonumber\\
&&\equiv \delta_{s,s^{\prime}}\alpha_{1}(M, m)(|\vec{p}|),
\end{eqnarray}           
\begin{eqnarray}
&&u(\vec{p},s,M)^{\dagger}v(-\vec{p},s^{\prime},m)\nonumber\\
&&=\frac{\xi(s)^{\dagger}\vec{\sigma}\cdot\vec{p}\xi(-s^{\prime})}{2\sqrt{EE^{\prime}}}[\sqrt{\frac{E^{\prime}+m}{E+M}}-\sqrt{\frac{E+M}{E^{\prime}+m}}]\nonumber\\
&&=\delta_{s,-s^{\prime}} \frac{(-1)^{(1-s^{\prime})/2}}{2\sqrt{EE^{\prime}}}[-\sqrt{(E^{\prime}+m)(E-M)}+\sqrt{(E+M)(E^{\prime}-m)}]\nonumber\\
&&\equiv \delta_{s,-s^{\prime}} (-1)^{(1-s^{\prime})/2}\alpha_{2}(M, m)(|\vec{p}|),
\end{eqnarray} 
\begin{eqnarray}
&&u(\vec{p},s,M)^{\dagger} \gamma_{5}u(\vec{p},s^{\prime},m)
\nonumber\\
&&=\delta_{s,s^{\prime}}\frac{(-1)^{(1-s^{\prime})/2}}{2\sqrt{EE^{\prime}}}[\sqrt{(E^{\prime}+m)(E-M)}+\sqrt{(E+M)(E^{\prime}-m)}]\nonumber\\
&&\equiv \delta_{s,s^{\prime}} (-1)^{(1-s^{\prime})/2}\beta_{2}(M, m)(|\vec{p}|),
\end{eqnarray}           
\begin{eqnarray}
&&u(\vec{p},s,M)^{\dagger}\gamma_{5}v(-\vec{p},s^{\prime}, m)\nonumber\\ 
&&=\delta_{s,-s^{\prime}}\frac{1}{2\sqrt{EE^{\prime}}}[\sqrt{(E+M)(E^{\prime}+m)} -\sqrt{(E-M)(E^{\prime}-m)}]\nonumber\\
&&\equiv \delta_{s,-s^{\prime}} \beta_{1}(M, m)(|\vec{p}|).
\end{eqnarray}

\end{document}